# A method to Implement the Kerberos User Authentication and the secured Internet Service

**Pak Song-Ho, Pak Myong-Suk, Jang Chung-Hyok**

**Kim Il Sung** University,

Pyongyang, DPR of Korea

**Abstract** This paper proposes a PKINIT_AS Kerberos V5 authentication system to use public key cryptography and a method to implement the gssapi_krb authentication method and secured Internet service using it in IPSec VPN

**Keywords** IPSec, VPN, Kerberos, Squid, GSSAPI

## 1. Introduction

Kerberos is a network authentication protocol created by MIT, and uses symmetric-key cryptography to authenticate users to network services. And IPSec is a typical security protocol that is used to construct VPN (virtual private network) in network layer of TCP/IP protocol.

When implementing secured Internet service combining Kerberos with IPsec security program, there are some problems.

① **Using public key cryptography in Kerberos authentication system**

Kerberos that was proposed as authentication system based on trusted third-party have been now developing by version 5[1] and it used public key cryptography to verify client's identification and share a secret key between user and Authentication Server or TGS server or Application Server[2, 4].

Specially, the method to use ECDH key exchange and ECSig signature in sharing secret key between the user and the AS server [5] is proposed, but program to implement these methods completely is not yet published.

② **Implementing gssapi_krb user authentication in IPSec**

Because the combination parameters between IPSec and Kerberos are generally set, the gssapi_krb authentication method[3,6,7] that the present IPSec proposes can be now not used. These parameters must be set according to application service name and mechanism type, etc.

③ **Implementing and using of secured Internet service in Kerberos-aware applications**

Httpd daemon with Kerberos authentication module(mod_auth_kerb) and Squid program with Kerberos authentication module(negotiate_auth) offer only a secured www service, and Kerberos V5 FTP program offers only secured ftp service of command-driven mode. (Kerberos-aware Email application is also same as Kerberos V5 FTP program)

In the former case, user must configure Firefox browser to use Kerberos for Single Sign-on. That is, it needs that user configure web browser to send user's own Kerberos credentials to the appropriate KDC. In the latter case, user don't use web browser.

If user wants to use Kerberos authentication service, user has to know how to use a web browser and secured FTP of command-based mode.

Paper proposes a PKINIT_AS Kerberos V5 authentication system to use public key cryptography and a method to implement the gssapi_krb authentication method and the secured Internet service using it in IPSec VPN.

## 2. Designing PKINIT_AS Kerberos Authentication System

Improved Kerberos authentication system using the public key cryptography is implemented on the assumption that KDC has public key certificates of clients previously, which has ① and ② message form other than default Kerberos authentication system.

① C-AS:Options,$ID_C$,$Realm_C$,$ID_{TGS}$,Times,Nonce1,**c_cert,{$ID_C$}$P_C^{-1}$**

② AS->C: $Realm_C$,$ID_C$,$Ticket_{TGS}$,{**{$K_{C,TGS}$}$P_C$** ,Times,Nonce1,$Realm_{TGS}$,$ID_{TGS}$}$K_C$

Where, Ticket$_{TGS}$ :{Flags,$K_{C,TGS}$,Realm$_C$,ID$_C$,AD$_C$,Times}$K_{TGS}$

③ C->TGS:Options,ID$_S$,Times,Nonce2,Ticket$_{TGS}$,Authenticator$_C$

Where, Authenticator$_C$ : {ID$_C$,Realm$_C$,TS1}$K_{C,TGS}$

④ TGS->C:Realm$_C$,ID$_C$,Ticket$_S$,{$K_{C,S}$,Times,Nonce2,Realm$_S$,ID$_S$} $K_{C,TGS}$

Where, Ticket$_S$ : {Flags,$K_{C,S}$,Realm$_C$,ID$_C$,AD$_C$,Times}$K_S$

⑤ C->S:Options,Ticket$_S$,Authenticator$_C$

Where, Authenticator$_C$: {ID$_C$,Realm$_C$,TS1}$K_{C,S}$

⑥ S->C :{TS2,Subkey,seq#}$K_{C,S}$

Where, Pc indicates public key of client C, $P_C^{-1}$ indicates private key of client C.

The authentication process of PKINIT_AS Kerberos authentication system proposed in this paper proceeds as follows.

① Client C sends signing its identification as own private key and its own public key certificate beside default request message to the AS.

② When AS receives the request message from client C and processes it, it identifies that public key of public key certificate in received message is the same as public key of public key certificate for the client in database, if these public keys are the same, it verifies signature using the key. If signature verification is successful, it proceeds with next step. AS sends encrypting secret session key($k_{c.tgs}$) between client C and Ticket Granting Server(TGS) with client's public key to the client.

③ Client decrypts received reply message from AS with secret Kc created by its password and then decrypts encrypted $k_{c.tgs}$ with its own private key $P_C^{-1}$ to get secrete session key $k_{c.tgs}$ between client and TGS.

The rest reception processes in client and the next steps of Kerberos authentication system are also same as default Kerberos processes.

## 3. Setting Combination Parameter to Establish the Security Contexts between the IPSec Client and the IPSec Server based on GSSAPI

To establish successfully the security contexts based on gssapi_krb between IPSec client and IPSec server, the Present IPSec security program sets combination parameter as follows.

**Step 1**: GSS-API initiator calls gss_import_name( ) to get the name of application server.

- Among parameters in this function. we set value of name type parameter to 《gss_nt_service_name》.

maj_stat = gss_import_name(&min_stat, &name_token, *(gss_OID) gss_nt_service_name*, &partner);

- Next, gss_canonicalize_name( ) is called to translate the above taken name to per-mechanism form. Among parameters in this function, we set value of mesh_type parameter to 《gss_mesh_krb5》.

maj_stat = gss_canonicalize_name(&min_stat, princ, *(gss_OID)gss_mech_krb5,* &canon_princ);

- From both these functions, server name to use in gss_init_sec _context( ) is gotten. Then, calling gss_acquire_cred ( ) function, we can acquire client's credential for use in gss_init_sec_context( ) function.

maj_stat = gss_acquire_cred(&min_stat, canon_princ, GSS_C_INDEFINITE, GSS_C_NULL_OID_SET, *GSS_C_ INITIATE* , &gps->gss_cred, NULL, NULL);

Where, 《*GSS_C_ INITIATE*》 is a value set in exchange for original value《GSS_C_BOTH》

**Step 2:** The client calls GSS_Init_sec_context() to establish a security context and sends the output_token to the server.

- .GSS_Init_sec_context() returns an output_token to be passed to the server
- The client may request various context-level functions through request flags of GSS_Init_sec_context().( for example, to enforce sequencing or to detect reply attack be applied to messages transferred on the established context and so on.)

The specification needs to pass GSS_C_MUTUAL_FLAG | GSS_C_REPLAY_FLAG | GSS_C_SEQUENCE_FLAG as req_flags, but original function hasn't GSS_C_REPLAY_FLAG req_flag.

We add GSS_C_REPLAY_FLAG req_flag as req_flags of GSS_Init_sec_context().

**Step 3**: Server receives the token from the client and transfer it to gss_accept_sec_context( ) and set it to value of INPUT_TOKEN parameter.

- A call to GSS_Accept_sec_context() at the server returns a token.
- Specially, before the server call gss_accept_sec_context(), it calls gss_import_name() and gss_canonicalize_name() to get client's name, and then, acquire server's credential calling gss_acquire_cred() with modified value as follow.

maj_stat = gss_acquire_cred(&min_stat, canon_princ, GSS_C_INDEFINITE, GSS_C_NULL_OID_SET, *GSS_C_ACCEPT,* &gps->gss_cred, NULL, NULL);

Where, *GSS_C_ACCEPT* is a modified value.

**Step 4:** The server creates OUTPUT_TOKEN and sends it to the client.

**Step 5:** The client and the server repeats and executes from step 2 to step4 till return value of gss_init_sec_context( ) and gss_accept_sec_context( ) sets to GSS_S_COMPLETE. If the return value is set to GSS_S_COMPLETE, a security context is set immediately.

**4. Implementing SSO User Authentication and the secured Internet Service**

① Using of IPSec transfer mode

To implement SSO user authentication and secured Internet service, this paper proposes security framework based on IPsec transfer mode.

This security framework consists of client (IPSec client program and Kerberos client program are running on a computer.), a server (IPSec) and a KDC authentication server, a www service(or a FTP or an E-mail service) can be run at the server.

First, user takes Kerberos ticket using PKINIT_AS Kerberos authentication system showed in the 2$^{nd}$.

Next, IPSec VPN with gssapi_krb authentication method using the method showed in the 3$^{rd}$. is constructed, as result of this, a security tunnel is formed between an IPSec client and an IPSec server.

The client requests Internet service to a appropriate application server using this security tunnel and takes given service. (The request and reply between the client and server are encrypted)

② Using of Security Gateway

When using above IPSec transfer mode, If there are lots of Internet requests about the same data, the amount of traffic is proposed.

On assumption that internal network with application servers is absolutely secure, this paper proposes security framework with IPSec, Kerberos and Squid proxy server.

This security framework consists of a security gateway program and a client program.

The client program includes a Squid client (IE or Firefox browser), a authentication client (Kerberos client) and an IPSec–VPN client and the security gateway program includes a Squid proxy server, a KDC authentication server and an IPSec–VPN server.

First, user takes Kerberos ticket using PKINIT_AS Kerberos authentication system showed in the 2$^{nd}$.

Next, IPSec VPN with gssapi_krb authentication method using the method showed in the 3$^{rd}$ is constructed, as result of this, a security tunnel is formed between an IPSec client and an IPSec server.

User requests Internet services to Squid proxy server using web browser.

## 5. Result analysis and Conclusion

When using a method proposed in the 2$^{nd}$ and ECC, security intensity was more increased $O(2^{n/2})$ ) than default Kerberos authentication system.

Because keys are stored at KDC, the total number of principal key is decreased as O(number of user × number of service)－O(number of user + number of service) than case that doesn't use a method proposed in the 2$^{nd}$.

If user uses a security framework based on the IPSec transfer mode and the security gateway, user doesn't need to know browser operation or command-driven mode, specially, using security policy database (SPD), user can request SSO user authentication and secured Internet service or not.

If the security gateway is used, because proxy service and the application layer firewall are provided, system performance and security performance can be raised.